# Indy: a virtual reality multi-player game for navigation skills training

Arnaud Mas*, Idriss Ismaël*, Nicolas Filliard*

* EDF R&D, Palaiseau, France

**ABSTRACT**

Working in complex industrial facilities requires spatial navigation skills that people build up with time and field experience. Training sessions consisting in guided tours help discover places but they are insufficient to become intimately familiar with their layout. They imply passive learning postures, are time-limited and can be experienced only once because of organization constraints and potential interferences with ongoing activities in the buildings. To overcome these limitations and improve the acquisition of navigation skills, we developed *Indy*, a virtual reality system consisting in a collaborative game of treasure hunting. It has several key advantages: it focuses learners' attention on navigation tasks, implies their active engagement and provides them with feedbacks on their achievements. Virtual reality makes it possible to multiply the number and duration of situations that learners can experience to better consolidate their skills. This paper discusses the main design principles and a typical usage scenario of *Indy*.

## 1 INTRODUCTION

Developing spatial understanding of an industrial building's structure can be quite difficult. Indeed, the facility priority use is production. In particular, learning sessions in such a building, even short ones, have to meet the production constraints such as busy schedule and safety measures.

Virtual learning environments (VLE) offer a great alternative for learning tasks difficult to undertake in the real world [4]. Indeed, they allow learners to visit a facility without any consideration about availability, distances or safety. They also give trainers the possibility to modify characteristics of the environment beyond what is possible in real life. They give learners the possibility to learn through active action instead of passive knowledge acquisition, as explained for instance by Pan et al. [14]. Besides, attention, active engagement, feedback and strengthening phases are important in a learning process to ensure effective acquisition of knowledge, as already identified by Dehaene [5] in the way children learn how to read.

*Indy* is a new virtual reality application for professionals who will work in industrial facilities. It aims at helping them get familiarized with the facilities during their training period. It is designed to provide professional trainers a tool to build new pedagogical strategies based on virtual reality.

## 2 DESIGN PRINCIPLES

### 2.1 Collaboration

Literature tend to show that fostering social interactions and collaboration leads to higher learning efficiency for a virtual group

* arnaud-a.mas@edf.fr, idriss.ismael@edf.fr, nicolas.filliard@edf.fr

[5][11][19]. Based on this hypothesis, we designed *Indy* to foster collaboration in a learning context, as defined by Roschelle et al. [16]: "the mutual engagement of participants in a coordinated effort to solve the problem together".

According to Slavin [19], group goals and individual accountability can contribute to collaborative learning achievements. Kreijns et al. [11] also cite positive interdependence and promotive interaction as levers to enforce collaboration. The authors insist on the fact that "the key to the efficacy of collaborative learning is social interaction, and lack of it is a factor causing the negative effectiveness of collaborative learning". Nonetheless, they point out that technology allowing communication won't automatically imply social interaction, and that off-task casual communication is important for group cohesion.

*Indy* offers trainers a tool to create a training scenario where several teams are immersed in a virtual industrial building, relying on an asymmetric collaboration method [12]:

- Some learners are immersed, using a head mounted display (HMD), in a virtual mockup of the building. They will play the "hunters", who will have to find their way to the objective.
- Some learners use floor maps and 360° photographs, on a desktop computer. They will play the "radios", who will have to guide the hunters.

With this method, learners have access to complementary information according to their role: each one has a key capability to achieve the scenario objective. This makes the communication within the team necessary to navigate in the virtual building. Besides, each team member has a particular point of view and professional background that can be shared with the other team members either to help one to be more efficient or to explain his/her choices.

In *Indy*, communication between teammates relies on two components: oral communication and pointing at objects in the 3D environment. This way they can help each other, share their viewpoints to confirm the itinerary (current position and future direction) and identify the objective.

However, communication is not limited to the VLE: *Indy* is designed as part of a full training sequence, during which the trainer and the learners are physically in the same room. Although VLEs allow virtual teams, with physically separated members [10], we wanted to preserve the training sequence, which fosters social interaction.

### 2.2 Gamification

Gamification can be defined as "the use of game design elements in non-game contexts" [6]. It is commonly used to increase users' motivation and engagement [8][15]. Nah et al. [13] list the following design elements commonly used in gamified applications, in the educational and learning contexts: points, levels/stages, badges, leaderboards, prizes and rewards, progress bars, storyline, and feedback. Sailer et al. [18] also list specific elements known to show positive effect on users' motivation: points, badges, leaderboards, performance graphs, meaningful stories, avatars and teammates.

*Indy* is a treasure hunt in an industrial facility: learners have to search in the building for a specific equipment or zone. These situations simulate the ones operators often face in real life, when planning an intervention in a building in which they can go only occasionally. This approach is similar to "investigation-scenarios" already experimented in learning sessions [7].

Trainers can launch a contest between teams: the time taken by the teams to find the objective is recorded and displayed at the end of the hunt. The trainer then has the opportunity to use the following elements:

- Level/stages: the trainer can create a new hunt in a few seconds, adapting the difficulty level to the class. The difficulty can vary according to the objective itself, but also to the presence of obstacles placed by the trainer, and the kind of information he/she gives orally or through the application before the hunt.
- Points / leaderboard: the trainer can use the recorded time of each team to that matter.
- Feedback: after the end of a hunt, an interface allows the trainer to show to the learners all their itineraries during the hunt and screenshots that he/she might have taken. He/she can use all this material for debriefing.

### 2.3 Spatial navigation training

3D VLE are an efficient tool for acquiring spatial navigation skills [4][1]. Waller et al. [20] suggest that using immersive VLE for spatial navigation training might be equivalent to training in a real environment. Indeed, Rodrigues et al. [17] confirmed that good spatial knowledge transfer can occur from virtual reality to the real world.

Chrastil et al. [3] affirmed that active navigation leads to better spatial knowledge. Particularly, active decision making plays a great role in elaborating a mental model of the place.

Carbonell-Carrera et al. [2] also insisted on the importance of maps as a complement to VLE for spatial orientation.

As a result, *Indy* aims at maximizing the spatial knowledge acquired during training:

- The "hunters" have to move along the entire itinerary, at a standard walking pace.
- The "radios" have to orally explain the itinerary to the hunters.
- Finding the objective involves decision making along the way, particularly when the trainer places obstacles.

Indy focuses on the development of 4 skills: navigating in the building (landmarks identification), reading a map (survey knowledge), converting information between both, and communicating route information to others.

### 3 BUILDING MOCKUP

*Indy* uses the virtual mockup of a reactor building. It includes a detailed 3D model reconstructed from laser scans, 360° photographs and updated floor maps [9].

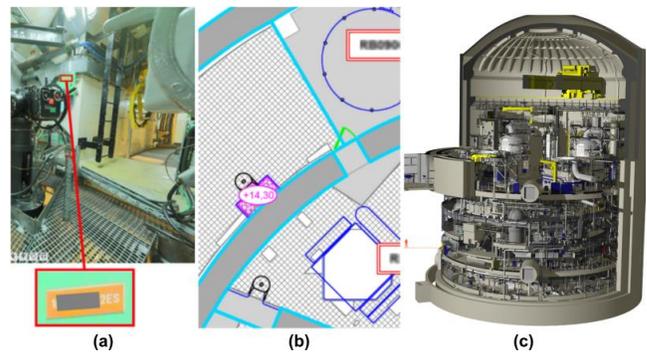

Figure 1: Overview of the elements of the building mockup used in *Indy*: (a) 360° photographs, (b) detailed floor maps, (c) 3D model (from Hullo et al. [9]).

### 4 A LEARNING SESSION WITH *INDY*

#### 4.1 Software overview

Indy was developed with Unity3D. It works as a network application, and provides three different clients depending on the users' role:

- Trainer (desktop): shows the floor maps and gives extended views and controls over the scenario.
- Learner (immersive): shows the 3D model in a HMD (HTC Vive).
- Learner (desktop): shows the floor maps, 360° photos and the 3D model. The learner can walk in the building using the mouse and the keyboard.

#### 4.2 Creating the learning scenario

When the trainer creates a new hunt, several configuration options are available:

- Type of hunt: the objective can be to point at a specific equipment or regroup into a target zone.
- Starting point: the trainer can choose any room of the building as a starting point.
- Objective: depending on the type of hunt, the trainer can either point at an equipment in the 3D model or draw a circle on a floor map.
- Objective as displayed to the learners: text that will be displayed, on demand, on the learners' side.
- Obstacles: the trainer can place obstacles or markings anywhere in the building.

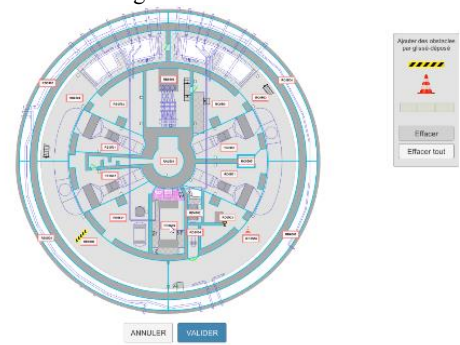

Figure 2: Interface allowing the trainer to place obstacles in the building.

### 4.3 Hunt

The learners are divided into teams of 3 or 4 members. In each team, one person plays the "radio" and the others play the "hunters".

Before starting the hunt, each team can gather around the radio's computer to prepare its itinerary. The learners can look for the equipment in the photos, and decide which itinerary seems to be best and even select alternative paths.

Afterwards, each hunter puts a HMD to start the hunt. All teammates can communicate orally, in order for the radio to give navigation instructions to the hunters. They also have the possibility to point at objects in the 3D environment, either with the mouse (for the radio) or with the Vive controllers (for the hunters). When doing so, a ray comes from the player's avatar and the object intersected by the ray is highlighted.

The players can't see players from other teams. The visibility between teammates is defined with the following rules:

- The radio can see the hunters' avatars (and pointing ray) in the 3D environment, and their positions on the floor maps.
- The hunters can see other hunters' avatars (and pointing ray), but not the radio's avatar (nor pointing ray). This keeps the hunters from simply following the radio.

During the whole hunt, the trainer can:

- See all the learners on the floor maps or in the 3D environment: he/she can choose which team(s) to see, and which team(s) can see his/her own avatar in the 3D environment.
- Point at objects (similarly to the radio players). The pointing ray will be seen by all players seeing his/her avatar.
- See all learners' point of view.
- Take screenshots: they are saved with their timestamp relative to the hunt.

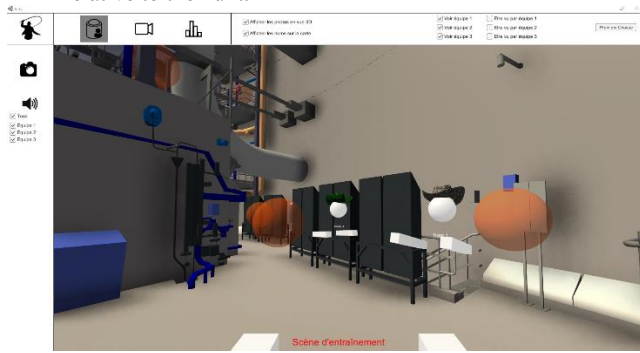

Figure 3: Trainer watching learners. He/she can either be visible or invisible.

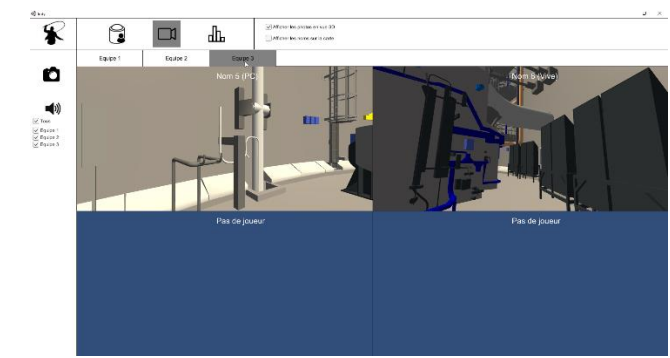

Figure 4: Trainer's interface showing different learners' point of view.

### 4.4 End of the hunt

To complete the hunt, all the hunters of the team must either (depending on the type of hunt):

- point at the target equipment at the same time: they have to keep pointing at the object for two seconds, while a validation animation is displayed ;
- be present in the target zone at the same time, and stay in it for two seconds (with the same validation animation).

Once the objective is found and validated, the team's total time is saved. When all teams found the objective, the hunt ends: on all screens, a score board shows the total time of each team. The trainer can then start the debriefing.

### 4.5 Debriefing

For a training session to produce better results, we propose feedback data to the trainer. On his/her computer, the trainer can see: the total time of each team in a single timeline, hunters' paths drawn on floor maps and screenshots he/she took during the session.

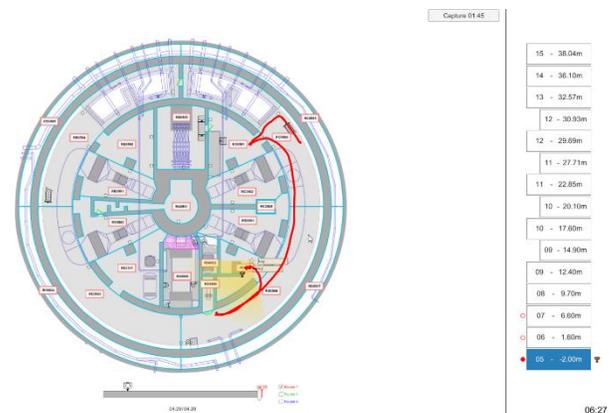

Figure 5: Debriefing interface, showing leaners' itineraries during the hunt, and the timeline on the bottom.

The debriefing interface is completely interactive. The trainer can move a cursor along the timeline to update hunters' paths on the maps. All the screenshots the trainer took are also situated on the timeline. The column on the right, used to change floors, also indicates with markers the floors where the learners currently are, synchronized with the timeline cursor.

The debriefing interface was designed as a tool, leaving the liberty to the trainer to use it as he/she wants, and display it in the classroom to focus on the elements pedagogically relevant for the training.

## 5 DISCUSSION

*Indy* is a tool for professional trainers, allowing them to create and propose scenarios adapted to all kinds of learners: beginners, experienced professionals, engineers or operators. The treasure hunts can be repeated and debriefed as many times as necessary to achieve the trainers' pedagogical objectives.

The debriefing is a key functionality for trainers. It is designed to be as little restrictive as possible, so that trainers would have the entire liberty to define the pedagogical valorization of the hunts. Among its possible uses, we can anticipate the following: discussing learners' itineraries and their implications in terms of safety, suggesting alternatives taking into account additional constraints than those experienced during the hunt, ask learners to explain their own choices, comparing different sessions with the same objective, or even evaluate long-term progression during the learners' careers.

The variety of scenarios that trainers can build with *Indy* offers several benefits. It allows trainers, when needed, to adapt to a specific work profile. Indeed, different professionals have different needs in terms of spatial knowledge of the facilities: emergency intervention teams, logistics supervisors, operators performing non-destructive testing, safety engineers, etc.

The obstacles adding functionality, in particular, plays a central role in adapting the scenarios. It also offers a way to evaluate the learners' global understanding of the building structure, as they will adapt their strategies in real time to find the objective.

By fostering collaboration and turning the learning experience into a game, we intend to focus learners' attention and actively engage them in using their navigational abilities. *Indy* offers feedback and allows trainers to integrate strengthening phases in the learning sequence. *Indy* is a tool and the learning scenarios built upon it are an expression of the pedagogical strategy trainers will choose. The best practices are yet to be explored.

However, *Indy* has yet to be experimented in a real learning session. In order to confirm our hypotheses on the pedagogical benefits of the proposed approach, a comprehensive study should be undertaken. Despite several promising studies, the empirical data is still sparse to fully understand the transfer mechanisms from virtual reality to real life, especially for professionals in an industrial context.

## 6 CONCLUSION

We presented *Indy*, a collaborative virtual reality application using gamification to help professionals acquire navigation skills needed to work in industrial facilities.

*Indy* consists in a treasure hunt in a virtual building, making learners cooperate in order to find an objective. It is designed to be integrated into existing training methods, and to be used by a trainer with a group of learners.

This paper summarized the design principles of *Indy* and its key functionalities offering pedagogical benefits.

## 7 ACKNOWLEDGMENTS

This work is funded by EDF R&D, and was developed by Florian Gavel during his 6 months internship.